\documentclass[amsmath,amssymb,reqno,tbtags,
twocolumn,showpacs,
pre, superscriptaddress]{revtex4}

\usepackage{upgreek} 
\usepackage{color,stmaryrd}
\usepackage{graphicx} 
\usepackage{hyperref}
\usepackage[capitalise]{cleveref} 
\usepackage{bm}

\newcommand{\av}[1]{{\ensuremath{\left\langle\vphantom{1^2} #1 
\right\rangle}}}
\newcommand{\dd}{\ensuremath{\text{d}}}

\newcommand{\fluid}{\ensuremath{_{F}}}
\newcommand{\matrx}{\ensuremath{_{M}}}

\newcommand{\self}{\ensuremath{_{\mathrm{s}}}}
\newcommand{\crit}{\ensuremath{^{\mathrm{c}}}}
\newcommand{\msd}{\ensuremath{\delta r^2}}
\newcommand{\mqd}{\ensuremath{\delta r^4}}
\newcommand{\ngp}{\ensuremath{\alpha_2}}

\newcommand{\order}[1]{\ensuremath{\mathcal{O}\left(#1\right)}}

\newcommand{\cut}{\ensuremath{_{\mathrm{cut}}}}
\newcommand{\thermal}{\ensuremath{_{\mathrm{th}}}}
\newcommand{\eps}{\varepsilon}

\begin{document}

\title[]{Dynamic heterogeneities and non-Gaussian behavior in two-dimensional \\
randomly confined colloidal fluids}

\author{Simon K. Schnyder}
\email{skschnyder@gmail.com}
\affiliation{Institut f\"{u}r Theoretische Physik II, 
Heinrich-Heine-Universit\"{a}t D\"{u}sseldorf, 
Universit\"{a}tsstra{\ss}e 1, D-40225 D\"{u}sseldorf, 
Germany}
\affiliation{Department of Chemical Engineering, 
Kyoto University, Kyoto 615-8510, Japan}

\author{Thomas O. E. Skinner}
\author{Alice L. Thorneywork}
\author{Dirk G. A. L. Aarts}
\affiliation{Department of Chemistry, Physical and Theoretical 
Chemistry Laboratory, University of Oxford, South Parks Road, 
Oxford OX1 3QZ, United Kingdom}

\author{J{\"u}rgen Horbach}
\affiliation{Institut f\"{u}r Theoretische Physik II, 
Heinrich-Heine-Universit\"{a}t D\"{u}sseldorf, 
Universit\"{a}tsstra{\ss}e 1, D-40225 D\"{u}sseldorf, 
Germany}
\email{horbach@thphy.uni-duesseldorf.de}

\author{Roel P. A. Dullens}
\email{roel.dullens@chem.ox.ac.uk}
\affiliation{Department of Chemistry, Physical and Theoretical 
Chemistry Laboratory, University of Oxford, South Parks Road, 
Oxford OX1 3QZ, United Kingdom}

\date{\today}



%
\begin{abstract}
A binary mixture of super-paramagnetic colloidal particles is confined between glass plates 
such that the large particles become fixed and provide a two-dimensional
disordered matrix for the still mobile small particles, which form
a fluid. By varying fluid and matrix area fractions and tuning the
interactions between the super-paramagnetic particles via an external
magnetic field, different regions of the state diagram are explored. 
The mobile particles exhibit delocalized dynamics at small matrix area fractions and localized motion
at high matrix area fractions, and the localization transition is rounded by the soft interactions [T. O. E. Skinner et al, Phys. Rev. Lett. {\bf 111}, 128301 (2013)]. Expanding on previous work, we find the dynamics of the tracers to be strongly heterogeneous and show that molecular dynamics simulations of an ideal gas confined in a fixed matrix exhibit similar behavior. The simulations show how these soft interactions make the dynamics more heterogenous compared to the disordered Lorentz gas and lead to strong non-Gaussian fluctuations.
\end{abstract}

\maketitle

\section{Introduction}

Slow relaxation phenomena are often linked to the appearance of
a diverging length scale.  While for the arrest of particles in
glass-forming fluids the relevance of a divergent length scale is
a highly controversial issue \cite{Cavagna2009,Berthier2011}, the
existence of such a length scale is obvious if the slowing down of the
relaxation dynamics is associated with an underlying continuous phase
transition \cite{Hohenberg1977}, such as, e.g., the critical
point of a liquid-gas transition \cite{Hansen2006} or a percolation transition \cite{Stauffer2003,Ben-Avraham2000}. A
paradigm for slow relaxation in combination with a percolation
transition is the Lorentz gas where a single tracer particle moves
through the free volume provided by an disordered matrix of obstacles
\cite{Hofling2013}. If the density of obstacles is sufficiently
high the tracer does not find any percolating path through the system
and is thus localized in a finite volume. At the percolation transition
of the free volume, where the tracer particle exhibits a transition
from a delocalized to a localized state, the tracer particle probes
the fractal structure of the free volume. This is associated with an
anomalous diffusion dynamics, as reflected in a sublinear growth of
the mean-squared displacement (MSD). Generalizations of the Lorentz model, for instance with many interacting particles, soft interaction potentials, or correlated matrix structures, have been investigated in both simulation~\cite{Kurzidim2009, Kurzidim2010, Kurzidim2011, Kim2009, Kim2010, Kim2011} and theory~\cite{Krakoviack2005, Krakoviack2007, Krakoviack2009}.

The original classical Lorentz-gas model \cite{Lorentz1905,Beijeren1982} assumes Newtonian dynamics
for the tracer particle and a hard-sphere potential for its interaction
with the obstacles. Here, the ``energy barriers'' that the tracer sees
when it travels through the arrangement of obstacles are infinitely
high. However, in a modified model with soft interactions between the
tracer and the obstacles this is no longer the case and the effective
barrier height provided by the obstacles depends on the energy of the
tracer particle. Thus, for a given obstacle configuration the effective
free volume that the tracer can explore is strongly correlated with
its energy. As shown in a series of molecular dynamics (MD) simulations
\cite{Schnyder2015}, in an ideal gas of tracer particles in a random
arrangement of soft obstacles each particle sees a different percolation
transition of the free volume according to the kinetic energy that has
been assigned initially to each of the tracer particles. As a consequence,
the self-diffusion coefficient, averaged over all the particles, does not
show a singularity but due to the heterogeneous dynamics of the tracer
particles it indicates a rounded transition. Only if an average over
tracer particles with the same energy is performed, a sharp transition
as in the hard-sphere Lorentz gas is seen. These results suggest that
the rounding of the transition is a generic feature of realistic,
soft systems.

Recently, we have presented an experimental realization of a
two-dimensional Lorentz-gas-like system \cite{Skinner2013}.
It consists of a binary mixture of super-paramagnetic colloidal particles confined between two
glass plates such that the larger colloidal particles are immobilized
and the smaller particles can move through the matrix formed by the
larger ones. In this experiment, the effective size of the particles
is varied by exposing the particles to an external magnetic field
that induces magnetic dipoles in the particles, leading to a repulsive
$r^{-3}$ interaction between them (here $r$ is the distance between two
particles).  By varying the strength of the external magnetic field,
the effective density of the matrix is changed while the structure of
the matrix remains unaffected. We have demonstrated that
the tracer particles, i.e.~the smaller particles, exhibit a transition
from a delocalized state at low effective matrix densities to a localized
state at high matrix densities \cite{Skinner2013}. This transition is expected to be rounded since
the energy of the Brownian particles is a fluctuating quantity and,
due to the soft $r^{-3}$ interaction with the obstacles, the barriers
seen by the tracers are not infinitely high as for hard interactions.
Since the apparent barrier heights depend on the energies 
of the tracer particles, each tracer sees a different matrix structure, which implies strong dynamical heterogeneities 
that have not been characterized to date.

Here, we discuss generic features of the structure and dynamics in heterogeneous media by comparing the results of colloidal experiments and MD simulations. First, we qualitatively characterize the tracer dynamics by calculating the single-particle probability distributions and discuss the structure of the matrix and
fluid particles in terms of the partial pair distribution functions. We then show that around the transition from a delocalized to a localized state, the dynamics of the tracer particles in both simulation and experiment exhibit strong dynamic heterogeneities that are associated with strong non-Gaussian fluctuations. To this end, we provide a detailed analysis of
simulation and experiment in terms of the self-part of the intermediate scattering function (SISF), $F\self(q,t)$ \cite{Hansen2006}, the mean-quartic displacement (MQD), $\mqd(t)$ \cite{Rahman1964}, and the non-Gaussian parameter (NGP), $\alpha_2(t)$ \cite{Nijboer1966,Boon1991}, thereby extending upon our previous work \cite{Skinner2013}. 
We find that a large fraction of particles can be already
localized while the MSD still appears diffusive. While this heterogeneity is typical
for the Lorentz gas, we find it to be enhanced when the artificial constraint of assigning the same energy to all particles in the simulation
is removed. As a consequence, the rounded delocalized-to-localized
transition of the tracer particles is associated with a strong increase
of $\alpha_2(t)$ on rather small and intermediate time scales, whereas
the $\alpha_2$ of the Lorentz gas indicates only small deviations from
Gaussian behavior. This strong increase of  $\alpha_2(t)$ is found in
the experiment as well.

\begin{figure*}
\includegraphics{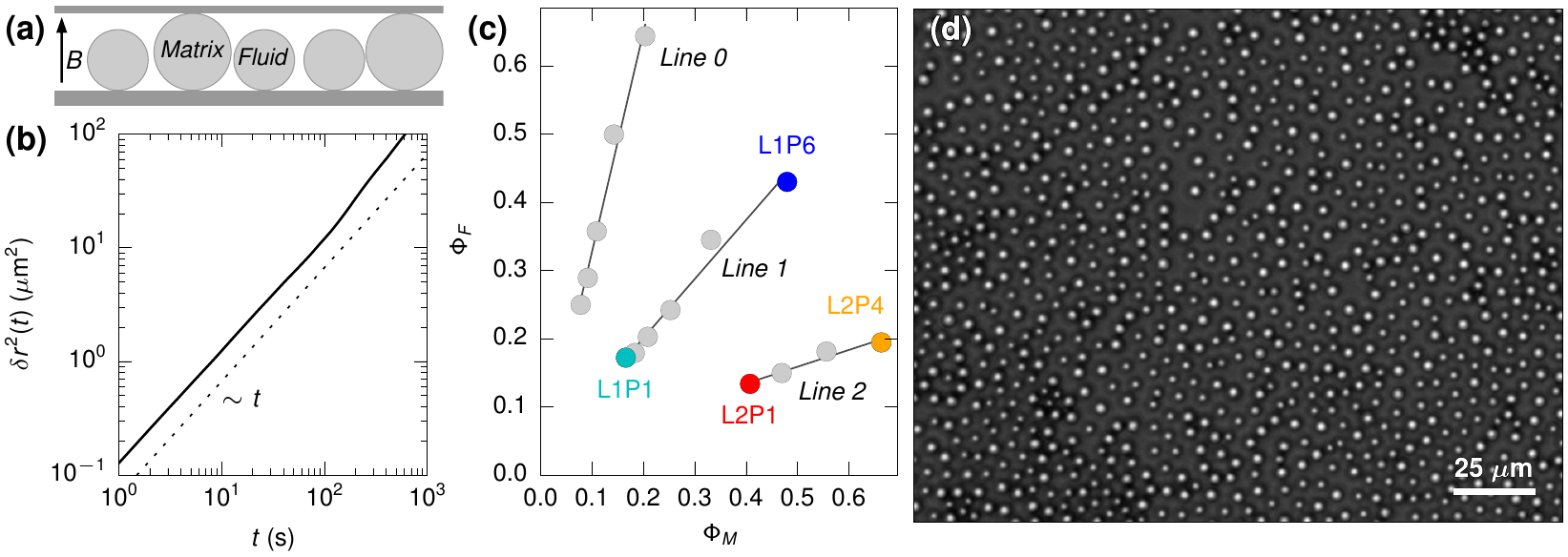}
\centering
\caption{a) Schematic of the experiment, a binary system of small and 
large particles confined between two glass slides (particle diameters to 
scale). The large particles support the top slide. The magnetic field $B$ 
tunes the effective interaction between the particles.
b) Snapshot of the system at state point L1P6 in a quadrant of size  $214 
\times 171 \upmu$m.
c) State diagram for the effective area fractions of the fluid 
($\phi_{F}$) versus the matrix particles ($\phi_{M}$).
d) Mean-squared displacement for the fluid particles in a very dilute 2D 
cell. A dashed line indicating diffusive behavior, $\msd(t)\sim t$, is 
shown as a guide to the eye. \label{fig:setup}}
\end{figure*}

\section{Colloidal model system}

The experimental system, as first introduced in \cite{Skinner2013}, consist of 
a binary mixture of super-paramagnetic polystyrene spheres (Microparticles GmbH) of diameters $\sigma_F^0 = 3.9~\upmu$m (index $F$ for 
\emph{fluid}) and $\sigma_M^0 = 4.95~\upmu$m ($M$ for \emph{matrix}), 
respectively, dispersed in water. The particles contain 
carboxyl surface groups that dissociate in water creating a short-range 
screened Coulombic repulsion. Their super-paramagnetic properties stem from the iron oxide nanoparticles 
distributed throughout their polymer matrix and a magnetic dipole will be induced parallel to an externally applied magnetic field. 

The binary colloidal suspension is confined between two glass slides to 
make a 2D sample cell. The large particles act as spacers to support the 
upper slide and form a fixed matrix, leaving the small particles --- the 
fluid --- free to move between them~\cite{Carbajal-Tinoco1997, 
CruzdeLeon1998, Santana-Solano2001, Santana-Solano2005}, see 
\cref{fig:setup}(a). To ensure that the small particles always stay in the 
plane, the height of the 2D sample cell, $h$, must be less than $h \approx 
1.447 \sigma_F^0$~\cite{Osterman2007}. The size ratio of the small to the 
large particles used in the binary mixture is selected accordingly and equals $0.787$. For the preparation of the sample cell, the lower and upper glass slides 
(Sail $76 \times 25 \times 1.2~$mm and Menzel-Glaser 15~mm $\times$ 15~mm $\times$
0.15~mm, respectively) are rinsed in distilled water, twice with absolute 
ethanol and then dried with an air gun. $1.11~\upmu$l of the required 
concentration of colloidal suspension is placed in the centre of the large 
glass slide to create a $15$~mm $\times 15$~mm $ \times 4.95~\upmu$m 
internal sample volume. The small glass slide is placed on top of the 
solution and a $10~$g weight pressure is applied to aid the liquid spread 
to the edges of the top slide. The edges of the sample cell are sealed 
with glue (Norland no. 82) and cured under a UV lamp. The cells typically 
last for 2 days before starting to dry out. 

After cell manufacture, the system is equilibrated for 30 minutes. The external magnetic field is set to the 
required value and the sample allowed to equilibrate for a further 20 
minutes. Using optical video microscopy stacks of 8-bit $1280 \times 1024$ 
pixel images of an area of size $428~\upmu$m $\times 342~\upmu$m are taken at 
typically 1~Hz for one hour. The colloidal particles are located by 
standard particle tracking routines \cite{Crocker1996}.  An optical 
microscopy image of the system is shown in \cref{fig:setup}(b). The 
colloids are fairly monodisperse, each with a coefficient of variation of 
$<3\%$, but this still leads to particles with sizes between the two. This 
slight size dispersity is noticeable from observation of the colloidal 
particles in the microscope image, \cref{fig:setup}(b).
Due to the size dispersity, sometimes fluid particles get stuck and matrix 
particles stay mobile. So particles are reclassified as fluid or matrix 
according to their mobility where required. This concerns only a very 
small fraction of the particles. Any drift in the colloidal particle 
positions in the microscope are corrected for with respect to the center 
of mass of the fixed 4.95~$\upmu$m particles. To improve statistics, each image is divided into quadrants. Each quadrant 
is analyzed separately and mapped onto the hard sphere state diagram. 
These data points are then binned according to their position on the state 
diagram to create points averaged over several similar matrix 
configurations and fluid particle densities.

The repulsive pair potential, $U_{F,M}(r)$, of the super-paramagnetic 
colloidal particles is controlled via an external magnetic field $B$:
\begin{align*}
	U_{F,M}(r) = \mu_{0} \chi_{F,M}^2 B^2/(4 \pi r^3),
\end{align*}
where $r$ is the distance between two particles, $\mu_0$ is the 
permeability of free space and $\chi_{F,M}$ the magnetic susceptibility of 
the fluid or matrix particles. For determining the effective packing 
fractions of the colloidal particles, effective hard sphere particle 
diameters $\sigma_{F,M}$ are calculated using the Barker-Henderson 
approach,~\cite{Barker1967, Henderson1977, Hansen2006}
\begin{gather*}
\sigma_{F,M} = \sigma_{F,M}^{0} + \int_{\sigma_{F,M}^0}^{\infty}~( 1 - e^{- \beta U_{F,M}(r)})dr,
\end{gather*}
where $\sigma_{F,M}^{0}$ are the hard sphere diameters of the colloids and 
$\beta = 1/k_{B}T$. If the magnetic field is switched off, $B = 0$, 
$\sigma_{F,M}$ reduces to the diameter of the colloids $
\sigma_{F,M}^0$, which corresponds to the lowest state point along each 
line. Hence, manipulation of both the number densities $n_F$ and 
$n_M$ of the colloidal particles, and the effective hard sphere diameters via the 
external magnetic fields allows different regions of the state diagram to be explored, see \cref{fig:setup}(c). 
We prepared three different samples with different number densities for 
the matrix and fluid particles, and thus investigated the system along 
three lines, labelled as lines 0, 1, and 2 in the state diagram.  The 
$n$-th state point of line $x$ is labelled as ``L$x$P$n$".  At the lowest 
point along each line the external field is not yet switched on and thus 
it is given by the hard sphere area fractions of the matrix $\Phi_M^0$ and 
the fluid particles $\Phi_F^0$. By switching on and increasing the 
magnetic field the effective area fractions are increased. The size ratio 
of the fluid to the matrix diameter stays constant at $0.787$, yielding 
linear paths in the state diagram.

The strength of this experiment lies in the fact that we are able to control 
the effective area fractions of the colloids without changing the matrix 
configuration. In this way, we can efficiently measure the tracer dynamics 
at a range of different effective matrix and fluid area fractions in the 
same sample. This approach allows us to achieve high matrix area fractions 
where the matrix still has a random character, a crucial property for a 
model system for random media. In our analysis of the tracer dynamics, we 
will focus on the state points along lines 1 and 2. The experimental data 
are averaged over up to four independent matrix configurations by imaging 
different parts of each sample.

In order to make sure that the dynamics of the colloids under confinement 
are well controlled, we prepared a system at a very low matrix packing fraction, with 
just enough particles to act as spacers, and containing a very low fluid 
particle concentration. With the 2D trajectory of any tracer particle 
designated as $\vec r(t)$, its mean-squared displacement (MSD) is defined 
by $\msd(t) := \langle (r(t)-r(0))^2 \rangle$, with $\left < \right>$ 
representing an average over different matrix configurations, i.e. 
multiple quadrants, employing multiple time origins, and averaging over 
all mobile particles. At such low packing fractions, the MSD
is expected to exhibit diffusion over all times, $\msd(t) \sim D_0 t$, with 
$D_0$ being the self-diffusion coefficient at infinite dilution, which is 
confirmed in \cref{fig:setup}(d). 
This indicates that the fluid particles are completely free to diffuse 
within the 2D cells. Note that diffusion is well-defined in 2D systems 
with obstacles \cite{Bauer2010}.

\section{Simulation}

In order to interpret the experiment, a molecular dynamics (MD) simulation of a comparable 
two-dimensional system was performed. Note that we are aiming to reveal qualitative and generic features of the localization dynamics 
across two quite different systems, rather than achieving quantitative agreement between experiment and simulation.
 
The fixed matrix in the simulation is generated from snapshots of an equilibrated 
polydisperse liquid of disks interacting with the Weeks-Chandler-Andersen 
potential \cite{Weeks1971}. The pair potential between particles is given 
by
\begin{align} 
V_{\alpha\beta}(r) = 
\begin{cases}
4\eps_{\alpha\beta} \left[
\left(
\left(\frac{\sigma_{\alpha\beta}}{r}\right)^{12}  -
\left(\frac{\sigma_{\alpha\beta}}{r}\right)^{6} \right)+ \frac{1}{4}
\right]
, & r<r\cut, \\ 
0, & r\geq r\cut, 
\end{cases} 
\label{eq:WCA} 
\end{align}
with a cutoff of $r\cut = 2^{1/6}\sigma_{\alpha\beta}$. The diameters of 
the matrix particles are sampled from an interval in order to avoid 
crystallization. The diameters of the $N$ particles are additive, i.e. 
$\sigma_{\alpha\beta} = (\sigma_\alpha + \sigma_{\beta})/2$, with 
$\sigma_{\alpha} = (0.85 + 0.3\alpha/N) \sigma_M$ and $\alpha,\beta = 
1,\ldots,N$. The unit of length is thus given by $\sigma_M$. The unit of 
energy is given by the energy scale for the matrix-matrix interaction 
$\eps_{MM}$. The numerical stability of the simulation is considerably 
improved by making the potential continuous at the cutoff. This is 
achieved by multiplying $V_{\alpha\beta}(r)$ with a smoothing function 
$\Psi(r) := (r-r\cut)^4/[h^4+(r-r\cut)^4]$ with width $h=0.005\, 
\sigma\matrx$. As a consequence, we do not observe any problems with 
energy drift in microcanonical simulations.
The particles are equilibrated with a simplified Andersen thermostat 
\cite{Andersen1980} at temperature $k_BT = \eps_{MM}$, where the particle 
velocities are randomly drawn every 100 time steps from the Maxwell 
distribution with thermal velocity $\text{v}\thermal := (k_BT/m)^{1/2}$. 
The unit of time is thus given by the Lennard-Jones time $t_0 := 
\sigma\matrx/\text{v}\thermal = [m(\sigma\matrx)^2/\eps\matrx]^{1/2}$. We 
integrate Newton's equations of motion for the particles with the 
velocity-Verlet algorithm~\cite{Binder2004} using a numerical timestep of 
$\Delta t = 7.2\cdot 10^{-4}t_0$.

We considered square-shaped systems containing $N = 500$, $1000$, $2000$, 
$4000$, and $16000$ particles, and employed periodic boundary conditions. 
To allow for sufficient averaging over different matrix configurations, we 
generated 100 statistically independent configurations for each case. 
These were equilibrated at the number density $n = N/L^2 = 
0.278/\sigma\matrx^2$, were fixed and subsequently uniformly rescaled to 
number density $n = 0.625/\sigma\matrx^2$, and thus correspond to the 
system sizes $L/\sigma\matrx = 28.28$, $40$, $80$, and $160$. Varying the 
system size $L$ allows us keep finite size effects under control.

Into the matrix structures, we insert a gas of tracer particles which do 
not interact with each other. The interaction of the tracers with the 
matrix is given by the WCA potential of \cref{eq:WCA} with parameters 
$\eps_{\alpha\beta} = 0.1 \eps\matrx$ and $\sigma_{\alpha\beta} = 
(\sigma\matrx + \sigma\fluid)/2$. Note that the polydispersity of the 
matrix particles is neglected here, as it was only used to avoid 
crystallization of the matrix configurations. The diameter $\sigma\fluid$ 
of the tracer particles acts as the control parameter and is used to 
change the area inaccessible to the tracer particles without changing the 
matrix structure, equivalently to modifying the magnetic field in the 
experiment. 

The tracer particles are inserted and equilibrated also with the 
simplified Andersen thermostat. Since the particles are non-interacting, 
the equilibration times can be quite short with run times of typically 
$10^3t_0$. For the microcanonical production runs we considered two cases. 
In the one case --- the confined ideal gas --- the production run is 
carried out directly after the equilibration, and the particles naturally 
have a broad distribution of energies. But the systems are first brought 
to the same average energy by rescaling all tracer velocities in each 
system with one constant, leaving the relative distribution of energies 
unmodified.  In the other case --- the single-energy case --- we enforce 
that all tracers have exactly the same energy. This is achieved by 
determining the average tracer energy at the end of the equilibration run, 
and reinserting the particles at random places, provided that their 
potential energy at that position is lower than the average energy and 
assigning the rest of the energy as kinetic energy. After that, 
microcanonical simulation runs are started for both cases with run times 
of up to about $2\cdot 10^5t_0$.
The single-energy case was shown to exhibit the universal critical 
behavior of the Lorentz gas with the transition occurring at the critical 
diameter $\sigma\fluid\crit \approx 0.43$, while the confined-ideal-gas 
case shows strong rounding \cite{Skinner2013,Schnyder2015}.

\begin{figure*}
\includegraphics*{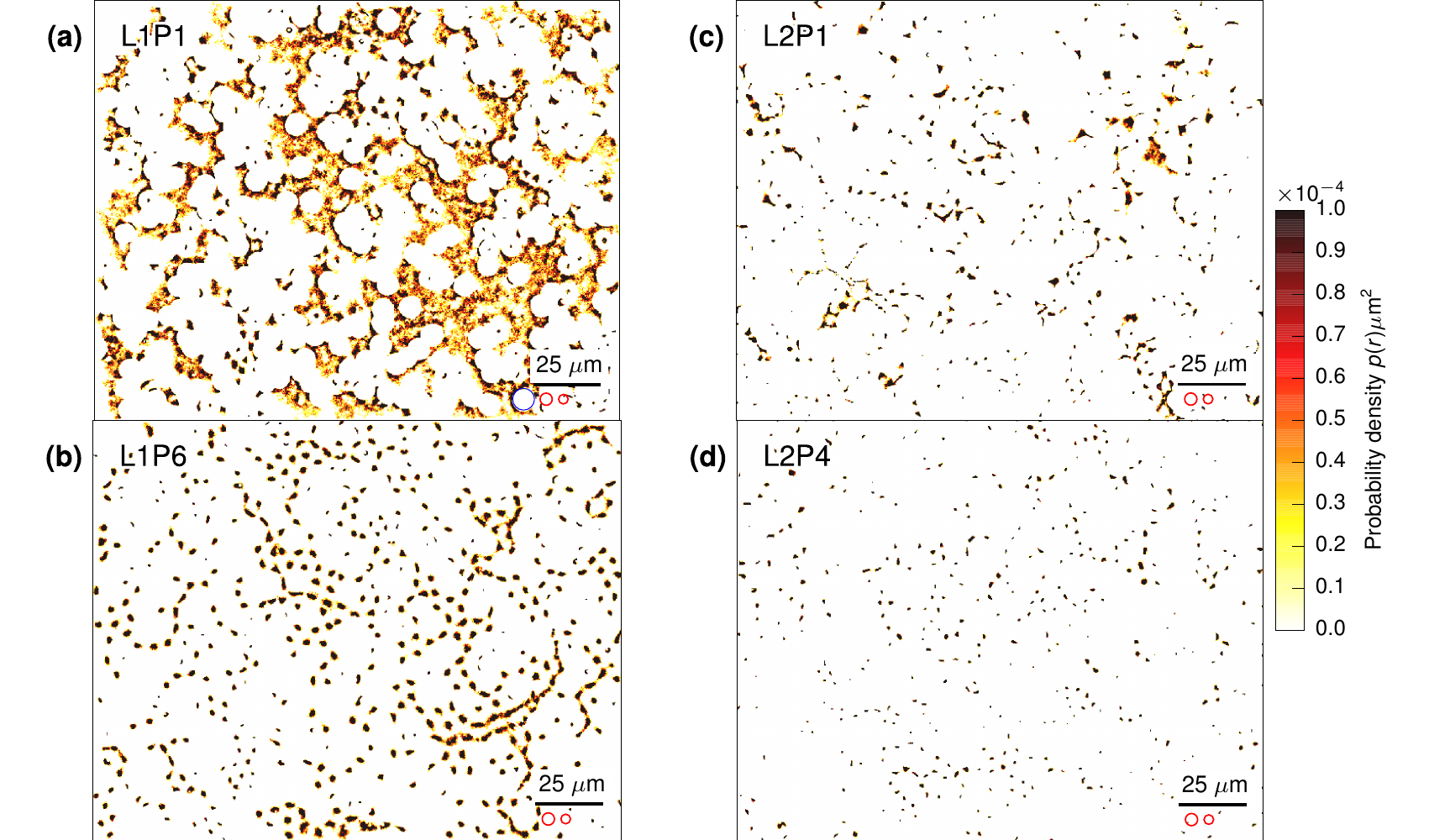}
\centering
\caption{Experiment: Single-particle probability distributions from 2D 
histograms of all colloidal particle positions in a quadrant measured over 
the duration of an hour for state points (a) L1P1, (b) L1P6, (c) L2P1, and 
(d) L2P6. Normalized such that the total probability of the whole quadrant 
is $1$. Size of the colloids annotated as red circles under the scales at 
the bottom right of each plot, size of the hard-core excluded area for 
centers of mobile particles indicated in (a) as blue circle under the 
scale.} \label{6hist2d}
\end{figure*}
\begin{figure*}
\centering
\includegraphics{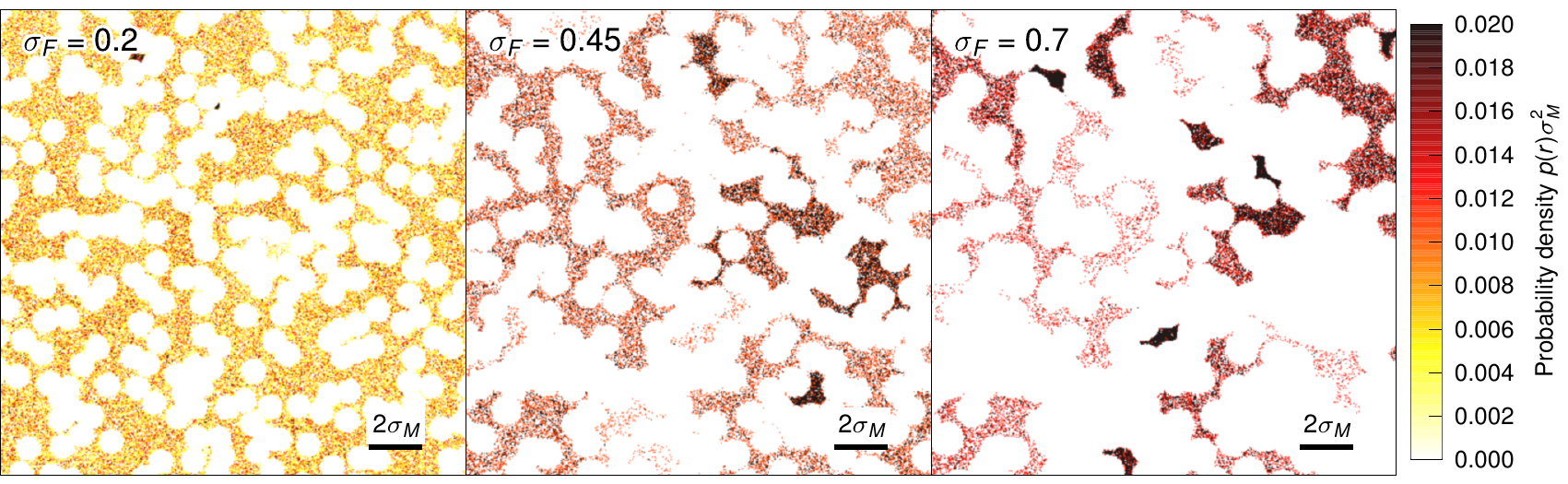}
\caption{Simulation: Single-particle probability distributions from 2D 
histograms in the confined ideal gas case. Shown are square sections of 
length $20\sigma\matrx$ and the histograms are normalized such that the 
total probability integrated over the shown section is $1$. 
\label{fig:histogram_simulation}}
\end{figure*}
\section{Results and discussion}
\label{6Results}
In our previous work~\cite{Skinner2013}, we used MD simulations to demonstrate that the experiment exhibits a 
delocalization-localization transition similar to the Lorentz gas, but that 
in contrast to the latter, the transition is rounded due to the soft 
interactions between the particles. 
Here, we revisit the experiment and the simulations, and analyse the 
structure of both the matrix and mobile component, as well as investigate 
the strongly heterogenous single-particle dynamics. With our analysis of 
the intermediate scattering function we get additional insights into the 
rounding of the localization transition, expanding on and complementing 
our previous work. As line 0 is very similar to line 1, it will be left 
out of the discussion.

\begin{figure*}
\includegraphics*{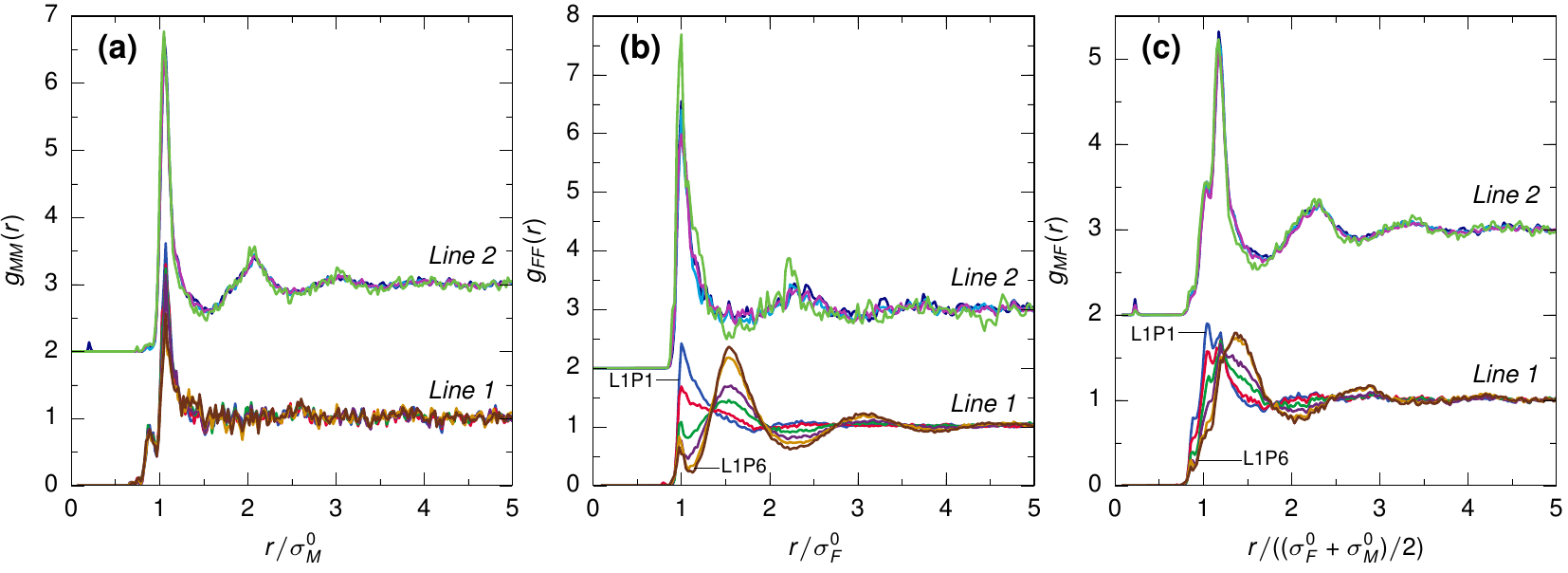}
\centering
\caption{Experiment: Partial radial distribution functions, for the (a) 
matrix-matrix interaction $g_{MM}(r)$, (b) fluid-fluid interaction 
$g_{FF}(r)$, and  (c) fluid-matrix interaction $g_{FF}(r)$ for each state 
point along lines 1 and line 2. Line 2 is shifted by 2 in each plot.
\label{fig:gr_experiment}}
\end{figure*}
\begin{figure*}
\includegraphics*{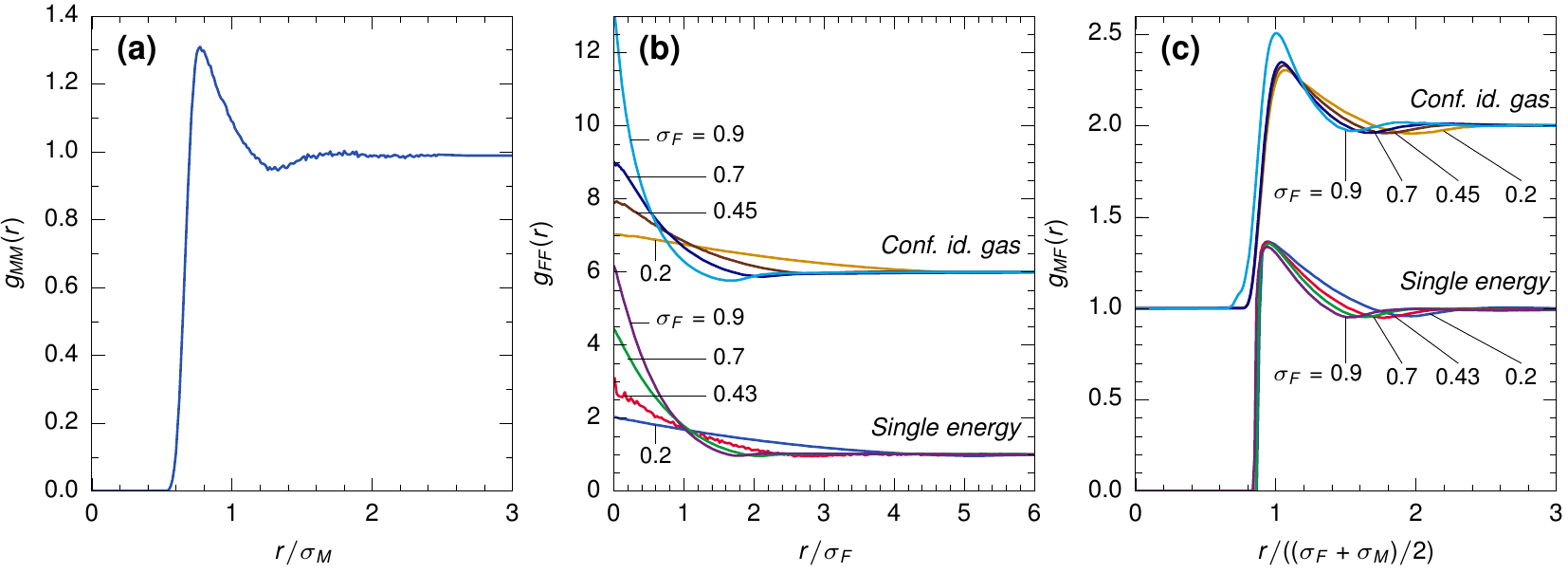}
\centering
\caption{Simulation: Partial radial distribution functions, for the (a) 
matrix-matrix interaction $g_{MM}(r)$, (b) fluid-fluid interaction 
$g_{FF}(r)$, and the fluid-matrix interaction $g_{FF}(r)$ for the 
single-energy and confined-ideal-gas cases (the latter data is shifted 
upwards by 5 in (b) and by 1 in (c)).} \label{fig:gr_simulation}
\end{figure*}
\subsection{Histograms}
Because we are able to track the full trajectories of the colloidal particles, we can directly
calculate the probability density $p(\vec r)$ of finding a single particle at position $\vec r$. To this end, we compute the histogram of all positions of the particle centers over 
the duration of the experiment (1h) on a grid where each bin corresponds 
to one pixel on the camera sensor, i.e. $\Delta A = (0.34\upmu 
\text{m})^2$, and normalizing the distribution such that the integral over 
a whole quadrant is unity. The distributions, shown in \cref{6hist2d}, give 
a good qualitative feel for the structure of the available free area and the dynamics of the tracer particles in 
the system and how it is modified when crossing the localization 
transition in the system. The obstacles are clearly visible as circular areas to which the fluid particles are excluded. At L1P1, where the magnetic field is switched off, the quadrant 
shown in \cref{6hist2d}(a) clearly shows a percolating path from the top 
center to the bottom right. 
At high magnetic fields, the motion of the fluid particles becomes severely 
constrained, see L1P6 in \cref{6hist2d}(b) where the same quadrant as in 
(a) is shown. The particles explore their surroundings, but on the time 
scales available to the experiment travel not much farther than their own 
diameter. 
This is not only due to the constriction of the matrix but also due to 
competition for free space between the mobile particles. However, the 
areas explored by the tracers are still connected in many cases, and large 
clusters of connected free area are found in the whole quadrant. It is 
probable that there is no percolating path present in the system and thus 
the sample is likely localized. Still, the MSD in 
this system becomes diffusive at long times \cite{Skinner2013}, which is an indicator for the 
rounding of the localization transition. The systems along line 2 are all strongly localized, regardless 
of the strength of the magnetic field, see \cref{6hist2d}~(c,d).

For qualitative comparison, we calculated analogous histograms for the 
simulation of the confined ideal gas, see \cref{fig:histogram_simulation}. 
The length scales are comparable, i.e. the obstacles are depicted at comparable 
size. 
At very small diameters, e.g. 
$\sigma\fluid = 0.2$ in \cref{fig:histogram_simulation}, the available 
area is highly connected, a situation that is not encountered in the 
experiment. The histogram at $\sigma\fluid = 0.45$ in 
\cref{fig:histogram_simulation} represents the situation close to the 
percolation point, where clusters of free area still span nearly the whole 
system. This is qualitatively comparable to the situation of L1P1 and 
L1P6. Highly dense systems contain only clusters of a linear extent of a 
few particle diameters, see $\sigma\fluid = 0.7$ in 
\cref{fig:histogram_simulation}. This situation is comparable to L2P1 and 
L2P4. While in certain ways the experiments and simulations are 
comparable, it is clear that it is extremely difficult to perform the 
experiments for long enough as to allow for the particles to sample the full 
available free area close to the critical point, a limitation that the 
simulations do not have.

\subsection{Matrix and fluid structure}
\label{6Dynamics}

In Ref.~\citenum{Skinner2013}, we characterized the structure of the 
matrix via the static structure factor and demonstrated that the matrix 
remains unchanged along each line, even at large magnetic fields $B$. It 
is revealing to also study the structure of the fluid, and as we have 
access to the trajectories of the particles in the sample, we can fully 
quantify the structural correlations in the system by calculating the 
partial radial distribution functions, 
$g_{\alpha\beta}(r)$~\cite{Hansen2006},
\begin{gather}
g_{\alpha\beta}(r) = \frac{A}{2\pi r}\frac{1}{f_{\alpha\beta}} \left 
\langle  \sum_{i=1}^{N_\alpha} \sum_{j=1, \neq i}^{N_\beta} \delta (r - 
|\vec{r_j} - \vec{r_i}| ) \right \rangle , \\ 
\nonumber
\text{with} \quad f_{\alpha\beta} = 
\begin{cases} 
N_{\alpha}(N_{\alpha}-1) 	&\text{ for }\alpha=\beta, \\
N_{\alpha}N_{\beta} 			&\text{ for }\alpha\neq\beta.
\end{cases}
\end{gather}
Here, $\alpha,\beta \in [F,M]$, $N_{\alpha}$ is the number of particles in 
component $\alpha$, $\vec{r}_i$ and $\vec{r}_j$ are the positions of 
particles $i$ and $j$ of components $\alpha$ and $\beta$, and $A$ is the 
area of the system or quadrant that is being evaluated.

In the experiment, the matrix-matrix component of the radial distribution 
function, $g_{MM}(r)$, for line 1, see \cref{fig:gr_experiment}(a), only 
exhibits a maximum for particles at contact, demonstrating that the matrix particles are nearly spatially uncorrelated. 
We observe a small pre-peak at $r \approx 0.9\sigma\matrx^0$ which 
probably comes from small particles getting stuck and thus being 
identified as fixed particles. The smallness of this peak demonstrates 
that this is only a very small effect. The function stays unchanged as the 
magnetic field is modified, demonstrating that the matrix particles really are 
fixed.
In contrast, the fluid structure as characterized by $g_{FF}(r)$ is strongly 
modified by the presence of the magnetic field, see 
\cref{fig:gr_experiment}(b). At zero magnetic field at L1P1 many 
particles are in contact, as demonstrated by the single maximum of 
$g_{FF}(r)$ at $r = \sigma\fluid^0$. With increasing magnetic field, the 
particles are driven further apart and the maximum decreases in amplitude. At the 
same time, another maximum appears and gradually shifts to larger 
$r$, in agreement with the growth of the effective diameter of the 
particles. Also, multiple smaller minima and maxima develop, indicating 
that the particles become more structurally correlated.
At L1P6 a small peak remains at the original position of the maximum ($ \sigma\fluid^0$), 
which indicates the a small portion of fluid particles cannot move away from 
each other even though the repulsive interaction is quite strong.
The matrix-fluid radial distribution function $g_{MF}(r)$ behaves quite 
similarly to $g_{FF}(r)$.

Line 2 differs from line 1 by having considerably larger number densities 
for both fluid and matrix particles. Consequently, the spatial 
correlations frozen in the matrix are stronger in line 2 as compared to 
line 1 and lead to a series of maxima and minima beside the main maximum 
of particles being at contact, see \cref{fig:gr_experiment}(a). Still, the 
matrix is fairly disordered with the extrema not being very pronounced. As 
for line 1, $g_{MM}(r)$ is independent of the magnetic field. In contrast to line 1, the fluid pair correlation function $g_{_{FF}}(r)$ is 
unchanged by the magnetic field, as well. This indicates that the 
particles are already so strongly confined by the matrix that increasing 
the repulsion between particles does not change their relative positions. The maximum 
of $g_{FF}(r)$ is very near the hard-sphere diameter of the particles, 
indicating that many fluid particles are at contact, fully occupying the free 
area inside the matrix and not leaving room to move around. Finally, similar to line 1, the matrix-fluid 
radial distribution function $g_{MF}(r)$ behaves quite similarly to 
$g_{FF}(r)$.

The data demonstrates the level of control we have over the structure of 
the system in the experiment. By varying the magnetic field, we can strongly influence the 
structure of the fluid, at least in the case of line 1 where the matrix 
density is moderate. The ability to calculate the partial pair correlation 
functions from the full trajectories of the colloidal particles demonstrates the 
strength of the colloidal model experiment, as the same would be very difficult to achieve 
in atomic systems or in analogous 3D colloidal systems with tuneable interactions.

In the simulation, the chosen matrix structure, see 
\cref{fig:gr_simulation}(a), is roughly comparable to 
the one found along line 1 in the experiment (\cref{fig:gr_experiment}(a)). 
Both are gas-like in structure with the experiment having a sharper peak. 
The main difference of the simulation to the experiment is that the 
simulated tracer particles do not interact with each other. This leads to 
considerably different structural correlations in the fluid, see 
\cref{fig:gr_simulation}(b). In contrast to the experiments, see 
\cref{fig:gr_experiment}(b), the particles are allowed to overlap, as indicated 
by the maximum of $g_{FF}(r)$ at $r=0$. As the particles become bigger, 
available space becomes increasingly rare, and the probability of tracers 
overlapping grows. Notably, the single-energy and confined-ideal-gas cases 
show very similar structural correlations. The matrix-fluid particle pair 
correlation function $g_{MF}(r)$ is also similar for both cases, see 
\cref{fig:gr_simulation}(c). The function exhibits a maximum at 
$r=(\sigma\matrx + \sigma\fluid)/2$, indicating that many tracers are at 
contact with matrix particles. The maximum of $g_{MF}$ in the confined 
ideal gas exhibits a less steep left shoulder due to the broad 
distribution of effective diameters in the system. As the size of the 
tracers increases, that maximum becomes sharper but stays at the same 
position. This is different from line 1 in the experiment and is again a 
result of the lack of interaction between the tracers.

\begin{figure}
\includegraphics{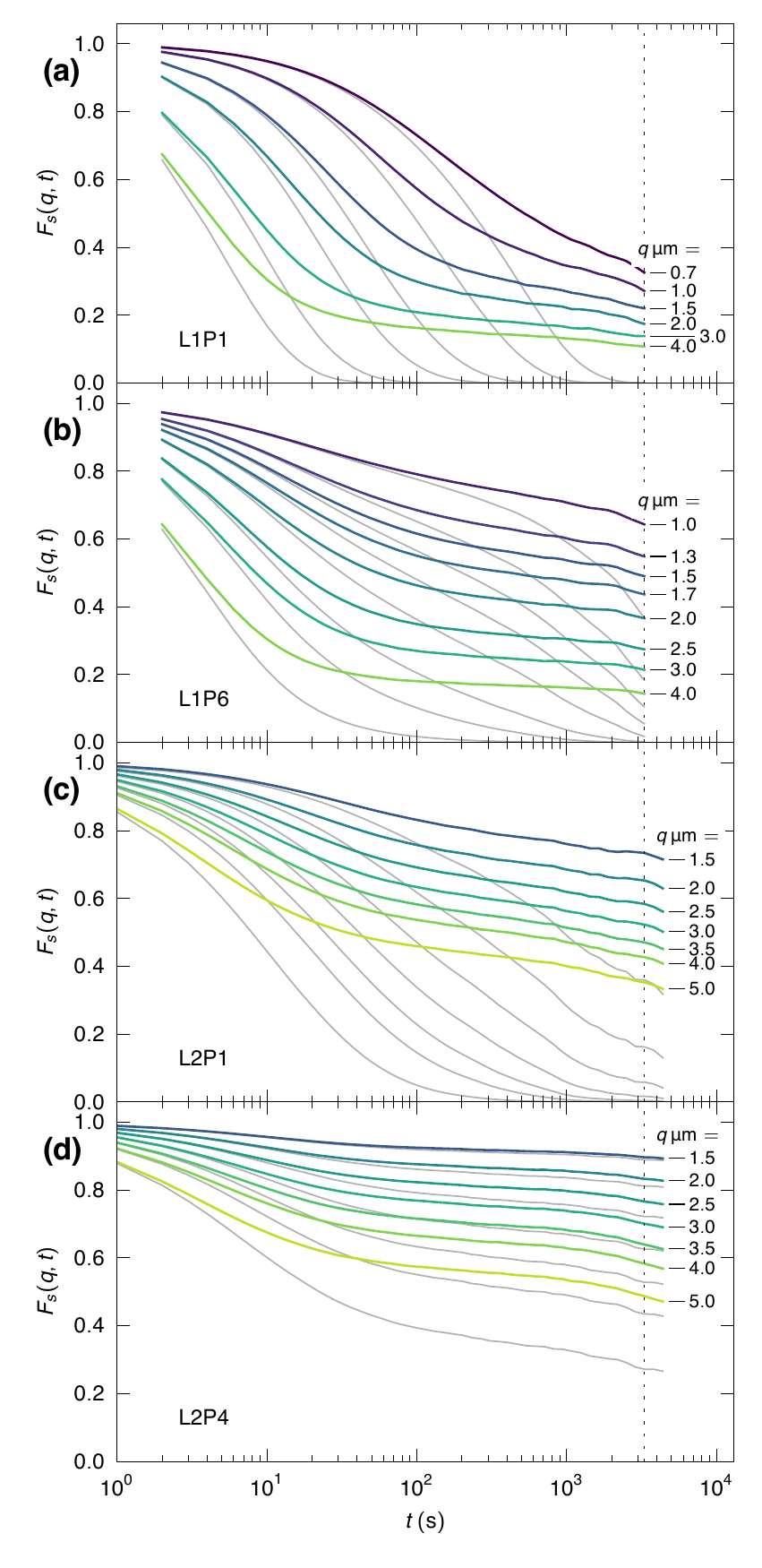}
\centering
\caption{Experiment: Self part of the intermediate scattering function 
$F\self(q,t)$ for the fluid particles for a range of wave numbers $q$ 
relating to state points (a) L1P1, (b) L1P6, (c) L2P1, and (d) L2P6 (in 
colors), as well as the corresponding Gaussian approximations (in grey). A 
measure of the non-ergodicity parameter is obtained with $f\self(q) 
\approx F\self(q,t\approx 3300 s)$, indicated by the dotted line, and 
shown in \cref{fig:fq}. \label{fig:fsqt_experiment}}
\end{figure}
\begin{figure}
 \includegraphics{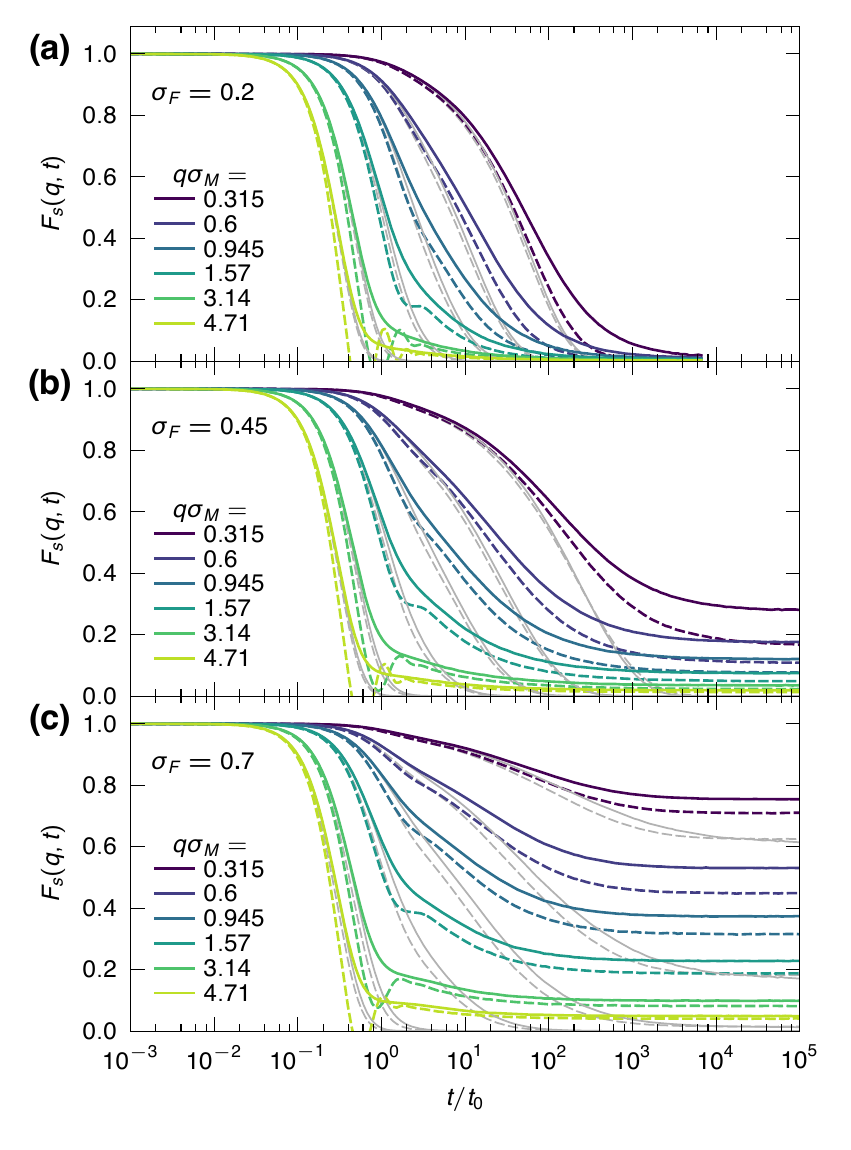}
 \caption{Simulation: Self part of the intermediate scattering functions 
in the simulation for the single-energy case (colored dashed lines) and 
the confined ideal gas case (colored solid lines) for particle diameters 
(a) $\sigma\fluid = 0.2$, (b) $0.45$, (c) $0.7$, as well as the 
corresponding Gaussian approximations (in grey). 
\label{fig:fsqt_simulation}}
\end{figure}

\subsection{Dynamics}
\subsubsection{Self-intermediate scattering function}

The self-part of the intermediate scattering function (SISF) for the 
mobile particles is defined as
\begin{equation}
F_{s}(q,t) = \frac{1}{N_F} \left \langle  \sum_{j=1}^{N_F} \exp\{i \vec{q} 
\cdot [\vec{r_{j}}(t) - \vec{r_{j}}(0) ] \} \right \rangle,
\label{SISF}
\end{equation}
with the 2D wave vector $\vec q$. Since the system is statistically 
isotropic, the SISF is invariant under the rotation of the direction of 
the wave vector and only depends on its magnitude, the wave number 
$q:=|\vec q|$. The SISF gives the full probabilistic information in 
Markovian systems and can be directly measured in scattering experiments. 
The SISF at any given $q$ describes the relaxation of density fluctuations 
on length scales $1/q$ over time. Its long-time limit $f\self(q):= 
\lim_{t\to\infty} F\self(q,t)$ is known as the non-ergodicity parameter or 
the Lamb-M\"o{\ss}bauer factor, and is a measure of the fraction of 
particles that are localized on a length scale $1/q$. Even though the 
self-part of the van Hove function discussed in Ref.~\cite{Skinner2013} 
contains the same information, it is of merit to study the SISF as well, 
since it is more sensitive to localized particles than both the van Hove 
function and its second moment, the mean-squared displacement, $\msd(t)$, 
which are more sensitive to highly mobile particles.

In the experiment, the SISF can be computed directly from the particle trajectories using Eq.~(\ref{SISF}) and we observe from \cref{fig:fsqt_experiment} that the 
SISF approximately has the same shape for all measured state points. The 
SISF decays in a single relaxation step onto a finite long-time limit 
$f\self(q)$, which increases with density, i.e. L1P1~$\rightarrow$~L1P6 and L2P1~$\rightarrow$~L2P4, and with larger length scales, i.e. smaller $q$. Note that this is qualitatively 
different from the two-step relaxation found in ideal glass formers~\cite{VanMegen1993,Gotze2008}. Even at the low densities 
of L1P1, fluid particles are trapped in voids created by the matrix, 
rendering the dynamics non-ergodic and preventing the SISF from decaying 
fully. For comparison, the SISF of the simulations are shown in 
\cref{fig:fsqt_simulation}.  Qualitatively, the single-energy and the 
confined ideal gas cases are extremely similar to each other. There is a 
single relaxation step onto a finite plateau which increases with 
increasing density, i.e. increasing $\sigma_F$, and larger length scales, i.e. smaller $q$. The main difference between the single-energy and confined ideal gas cases can be 
found in the short-time behavior around $t/t_0=\order{1}$, where the 
single-energy case resolves the first collision of the tracers, while this is 
averaged out in the confined-ideal-gas case. Apart from that, only the 
magnitude of the long-time limits is different in the two cases. In 
extremely dilute systems, e.~g. $\sigma\fluid = 0.2$, the long-time limit 
is nearly 0, indicating that only a small fraction of particles is 
localized. At larger diameters, e.~g. $\sigma\fluid = 0.45$ and $0.7$ in 
\cref{fig:fsqt_simulation}~(b, c), the long-time limits are finite, and 
the SISF of experiment and the confined-ideal-gas case in the simulation 
become qualitatively very similar.

\begin{figure}
\includegraphics[width=\columnwidth]{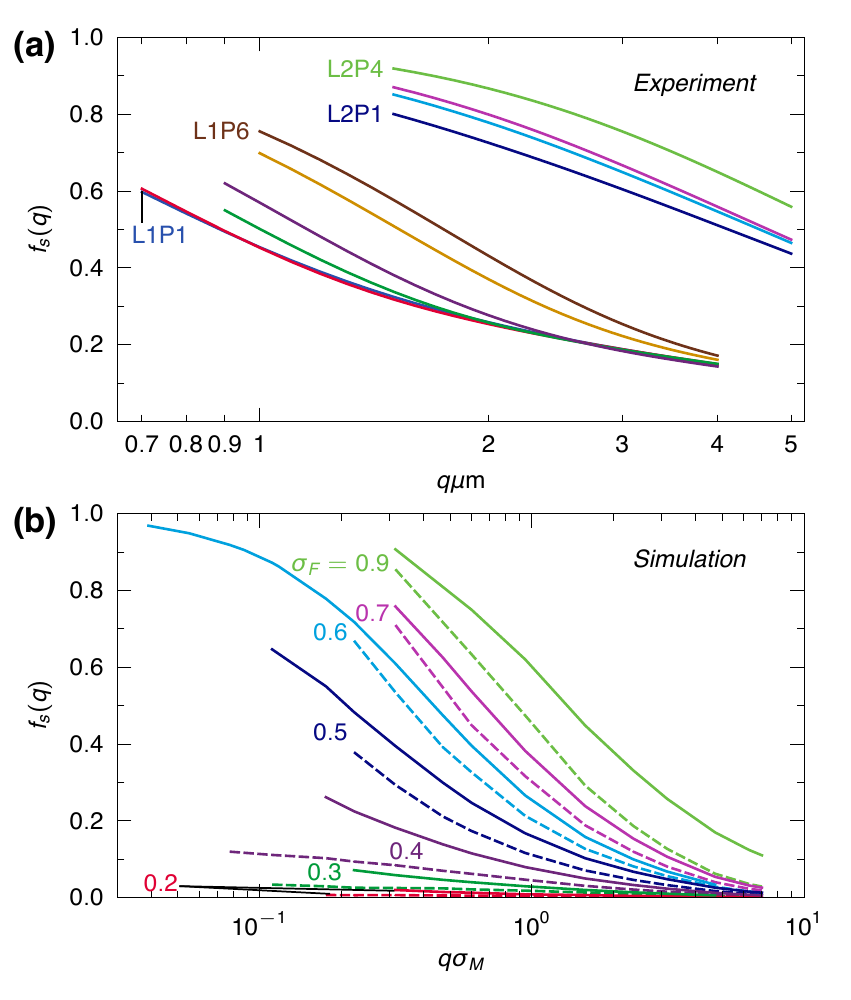}
\caption{(a) Long-time limit of the SISF $f\self(q)$ of the experiment 
along lines 1 and 2 in semilogarithmic presentation. 
(b) Long-time limit of the SISF $f\self(q)$ in the simulation
for the single-energy (dashed) and ideal-gas case (solid).} \label{fig:fq}
\end{figure}
To quantify the proportion of localized particles in the experiment, we 
approximately determined $f\self(q)$ as the value of $F_{s}(q,t)$ at $t 
\approx 3300\,\text{s}$  (indicated by the vertical dashed lines in 
\cref{fig:fsqt_simulation}) for all points along both line 1 and 2, see 
\cref{fig:fq}~(a). Note that this simply corresponds to the longest accessible timescale in the experimental data. The $f\self(q)$ for the simulations, shown in 
\cref{fig:fq}~(b), are easy to obtain as the simulations have shorter 
relaxation time scales. 
Qualitatively, the $f\self(q)$ of the experiment and simulations exhibit similar dependence on $q$. 
The $f\self(q)$ of line 1 of the experiment is similar to the $f\self(q)$ of the simulation for small $\sigma\fluid$ and the $f\self(q)$ of line 2  corresponds to that of the simulations at large $\sigma\fluid$. In all experimental state points $f\self(q)$ is finite, showing that even 
at the lowest densities along each line there are subsets of particles 
that are localized, similarly to the Lorentz model. Importantly, the SISF 
and $f\self(q)$ of both experiment and simulation look qualitatively the 
same on both sides of the transition. 

From the simulations we can further conclude that the dynamics is more 
heterogeneous in the confined ideal gas than in the single energy case, 
i.e.~the Lorentz gas. This is inferred from the fact that the $f\self(q)$ at the same 
$\sigma\fluid$ is larger in the confined ideal gas case, indicating a larger fraction of particles is localized, while, at the same time, the MSD of the confined ideal gas grows faster at long times that that of the single energy case (see Fig. 2b in Ref.~\citenum{Skinner2013}), indicative of more highly mobile particles. This increase in heterogeneous dynamics in the confined ideal gas case as compared to the single energy case is a trivial consequence of the broad energy 
distribution of the particles.

\subsubsection{The Gaussian approximation}

Next, we analyse the cumulants of the SISF, since this exposes dynamical heterogeneities 
more clearly. The SISF can be expressed via a cumulant expansion for small 
wave numbers as \cite{Hofling2013} 
\begin{align*}
	F_s(q,t) = \exp\left[
		- \frac{q^2 \msd(t)}{4} + \frac{1}{2} \alpha_2(t) 
\left(\frac{q^2 \msd(t)}{4}\right)^2 + ...
	\right],
\end{align*}
with the non-Gaussian parameter (NGP), $\alpha_2(t)$, relating the MSD, $\msd(t)$, and the 
mean-quartic displacement (MQD), $\mqd(t)$, to each other \cite{Rahman1964}:
\begin{align*}
	\ngp(t) := \frac{1}{2}\ \frac{\mqd(t)}{[\msd(t)]^2} - 1.
\end{align*}
The cumulants $\delta r^n (t)$ are defined as 
\begin{align}
	\delta r^n(t) := \av{|r(t) - r(0)|^n} = \int |r|^n P(r,t)\ \dd^d r,
	\label{eq:cumulants}
\end{align}
with the self-van-Hove function $P(r,t)$ being the one-particle density 
autocorrelation function in space and time, and the inverse Fourier transform of the SISF,
\begin{align*}
	P(r,t) := \frac{1}{N_F} \av{ \sum_{j=1}^{N_F} \delta\left[r - 
\left|\vec r(t) - \vec r(0)\right|\right]}.
\end{align*}
The odd-numbered cumulants vanish due to the rotational symmetry of the 
system. 

If the system exhibits diffusion, 
the SISF is Gaussian at all times and can be directly related to 
the MSD as follows
\begin{align}
	F\self(q,t) = \exp\left(- \frac{q^2\msd(t)}{4}\right),
	\label{eq:gaussian_approx}
\end{align}
which is known as the Gaussian approximation \cite{Hansen2006}. This approximation is valid for many systems, e.g.~for the diffusive motion of hard spheres \cite{Thorneywork2016}, or for harmonic oscillators \cite{Vineyard1958}, and $\alpha_2(t)$ is either 
exactly or close to 0 in these cases. 
The failure of the Gaussian approximation indicates, for instance, the presence of correlated motion, 
localization of particles on many different length scales, or the presence 
of multiple relaxation times, and is a strong indication of 
dynamical heterogeneity in the system~\cite{Kob1997a,Yamamoto1998,Weeks2000}, which is often quantified using the NGP.
In the Lorentz model the Gaussian approximation fails as well \cite{Lowe1997}, in particular at the critical point and 
$\alpha_2(t)$ never decays to 0, but instead exhibits a 
divergence~\cite{Spanner2011}. This is a result of the particles being 
confined in a fractal structure and leads to its extended subdiffusion.  

We find that the Gaussian approximation provides a good description of the SISF of all experimental and simulation data at short times, see 
\cref{fig:fsqt_experiment,fig:fsqt_simulation}. At long times it typically 
fails to capture the long relaxation times and the plateau heights. If the 
system becomes extremely localized, however, the Gaussian approximation matches the 
SISF more closely again, as illustrated in \cref{fig:fsqt_experiment}(d) for the experimental data at L2P4. Here, the particles mostly vibrate in small 
cages created by both matrix and neighboring fluid particles and the dynamics then effectively approaches the idealization of localization in harmonic 
potentials. As a consequence, the Gaussian approximation is found to be least successful close to the localization transition, as expected for the Lorentz model.

\begin{figure}
  \centering
    \includegraphics{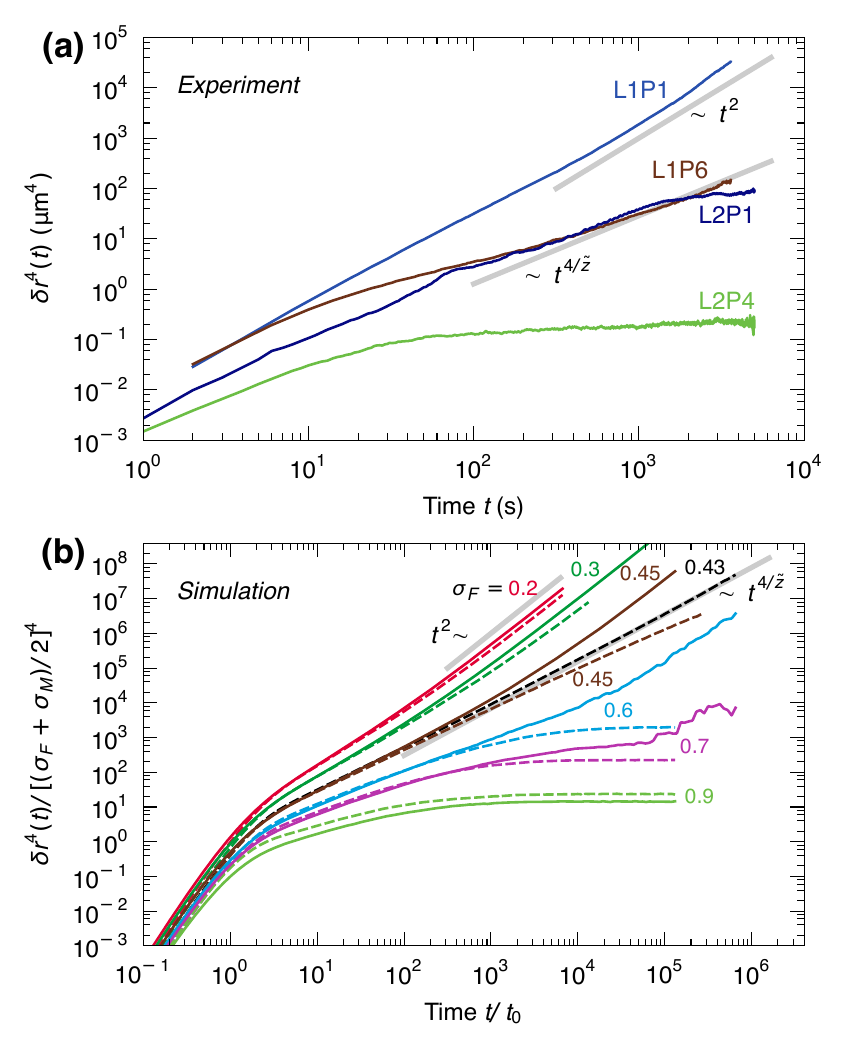}
  \caption{Mean-quartic displacement $\mqd(t)$ for experiment (a) and 
simulation (b) as a function of time. In the simulation, single-energy 
case (dashed lines) and confined-ideal-gas case (solid lines) are shown. 
The straight grey lines $\sim t^2$ and $\sim t^{4/\hat z}$ with $\hat z \approx 2.955$ 
serve as guide to the eye. \label{fig:mqd}}
\end{figure}

\subsubsection{The mean quartic displacement}

In the Lorentz model the mean-quartic displacement is expected to grow as $\mqd(t) \sim t^2$ at 
long times in the delocalized state, corresponding to regular diffusion, and becomes 
constant in the localized state. Close to the 
transition, it is expected to grow as $\mqd(t) \sim t^{4/\tilde z}$ with 
exponent 
 $\tilde z \approx 2.955$ in two dimensions \cite{Hofling2006}. The experimental data exhibit a transition from delocalized dynamics 
at L1P1 to localized dynamics at L2P4, with subdiffusive growth of the MQD 
at L1P6 and L2P1. The growth of the MQD at L1P6 at large times seems very loosely 
compatible with the Lorentz-model power law at the transition, but at closer 
inspection has a lower effective exponent. The simulation in the 
single-energy case is in full agreement with the Lorentz-model scenario, see 
\cref{fig:mqd}(b), making the transition from delocalized to localized 
dynamics and exhibiting extended power-law growth at the critical point at 
$\sigma_F = 0.43$ with the expected exponent. This shows once more that 
the single-energy case falls in the same universality class as the Lorentz 
model. The MQD for confined ideal gas shows strong rounding similar to the MSD 
\cite{Skinner2013}: the MQD exhibits the transition from delocalized to 
localized behavior but the transition is smoothed due to the averaging 
over a wide range of particle energies, which results in a wide range of effective exponents rather than the critical 
asymptote. Strikingly, at $\sigma_F = 0.6$, the MQD of the confined ideal gas initially follows the corresponding curve of the 
single-energy case, indicating localization of most particles, but at long 
times becomes dominated by the contributions of a few highly mobile, 
delocalized particles. This leads to subdiffusion over many orders of 
magnitude in time with an effective exponent smaller than the critical 
one -- similar to what is observed for the MQD in the experiment at L1P6 -- before crossing over to $\sim t^2$ at long times in the simulation. Note that we do not reach this time scale in the experiment. All of this is characteristic of the rounding of the localization transition. 

\begin{figure}
  \centering
		\includegraphics{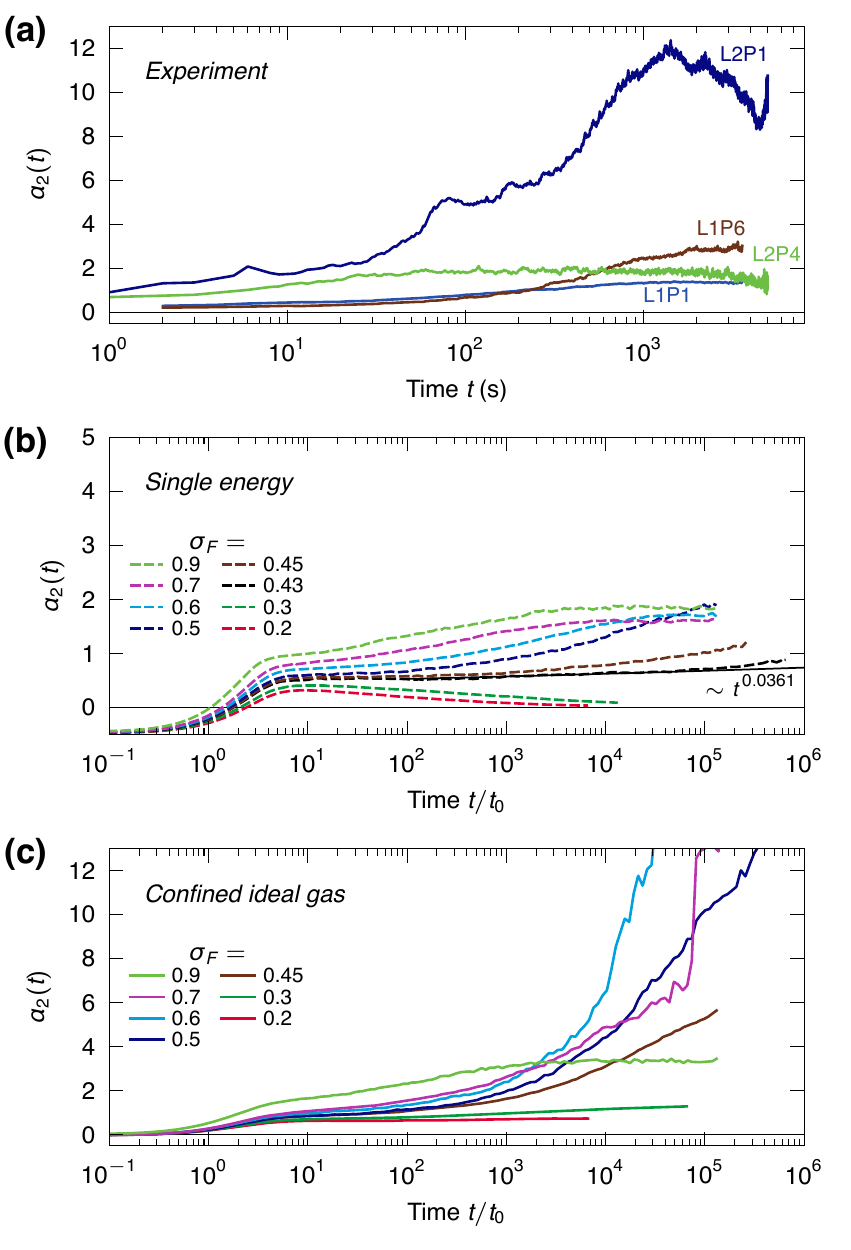}
  \caption{Non-Gaussian parameter for the experiment (a), and for the 
simulation in the (b) single-energy and (c) confined-ideal-gas
cases. \label{fig:ngp}}
\end{figure}

\subsubsection{The non-Gaussian paramater}

The non-Gaussian parameter (NGP), $\alpha_2(t)$, is very sensitive to dynamical heterogeneities \cite{Kegel2000, Shell2005}. In the experiment, the NGP on the delocalized side of the transition, i.e. along line 1, grows from nearly zero, characteristic of regular diffusion at short time scales, to values around 2 at long times for both state points L1P1 and L1P6, see \cref{fig:ngp} (a). On the timescale of the experiment, these NGPs do not decay, clearly showing that the dynamics remains non-Gaussian and heterogeneous. Note that this is qualitatively different from typical glassy dynamics, where the NGP goes through a maximum at intermediate times and decays to zero at long times \cite{Kim2011}. Along line 2 of the experiment, i.e. on the localized side of the transition, the NGP is already close to unity at early times for both L2P1 and L2P4. While L2P4 remains relatively constant but finite, the NGP for L2P1 -- which is close to the localization transition -- grows strongly with time. 

To interpret the behaviour of the NGP in the experiment, we now discuss the NGP in the simulation. First we consider the single energy case, which reproduces the Lorentz model~\cite{Hofling2006,Hofling2007,Hofling2008,Spanner2011}. In this case, the NGP parameter exhibits critical divergence at the localization transition, $\sigma_F = 0.43$, see \cref{fig:ngp} (b). Indeed, the exponent is nearly indistinguishable from the expected critical exponent of $\approx 0.0361$ in two dimensions  \cite{Hofling2006}. The small deviation from the asymptote at long times is most likely due to lacking statistics although we cannot fully rule out small finite size effects, which have been shown to particularly affect the NGP~\cite{Spanner2011}. However, the experiment and the single energy case clearly exhibit very different behavior and the critical divergence of the Lorentz model is so small that it cannot explain the experimental data. 

Therefore, we now consider the NGP of the confined ideal gas, which is shown in \cref{fig:ngp} (c). In this case, the NGP grows monotonically to long time values that are generally larger than those found in the single-energy case (note the different scales of the axes). Close to the rounded localization transition, $0.45 \leq \sigma_F \leq 0.7$, the NGP exhibits very strong growth, far exceeding those of the single-energy case. 
Strikingly, the confined ideal gas shows qualitatively similar behavior to the experiment, while being very different from the Lorentz model scenario seen in the single energy case. Importantly, this indicates that the observed heterogeneous and non-Gaussian dynamics in the experiment are not due to critical dynamics, but are a direct result of the rounding of the localization transition. In other words, the divergence of the NGP in the confined ideal gas is different from the weak critical divergence of the NGP in the Lorentz 
model. Because the NGP is not very sensitive to the critical dynamics, it exposes the non-Gaussian dynamics that occurs in the experiment and the confined ideal gas due to the rounding of the localization transition.

\section{Conclusion}
We have studied the dynamics of a quasi-two-dimensional colloidal fluid confined in a strongly heterogeneous matrix. The experiment exhibits a rounded localization-delocalization transition, in which the critical point is seemingly avoided. We have shown that the dynamics in the experiment is strongly non-Gaussian and by comparing the experiment to molecular dynamics simulations of a confined ideal gas, we have demonstrated that the heterogeneous and non-Gaussian dynamics are a generic feature of the \emph{rounding} of the localization transition. In addition, we have characterized the structure of the confining matrix and fluid particles in terms of the partial pair distribution functions.

The anomalous dynamics close to the transition has been analyzed with a particular focus on dynamical heterogeneities, by consideration of the self part of the intermediate scattering function, the mean-quartic displacement, and the non-Gaussian parameter. The self intermediate scattering functions decay in one step to their long-time limit, similar to the Lorentz gas, but different from typical glassy behavior. A large fraction of particles can be already localized while the mean-squared displacements, discussed in Ref.~\citenum{Skinner2013}, is still diffusive. Although this heterogeneity is typical for the Lorentz gas -- which is reproduced in our simulations when all the particles are assigned the same energy -- we have found that the heterogeneity is significantly enhanced when this energy constraint is removed and a confined ideal gas is considered. Strikingly, this leads to a strong increase of the non-Gaussian parameter close to the rounded localization transition, as also found in the experiments, which is different from the weak divergence predicted for the Lorentz gas. The comparison between the experiment and the simulations show how the soft interactions make the dynamics more heterogenous compared to the Lorentz gas and lead to strong non-Gaussian fluctuations.

\begin{acknowledgements}
We thank Felix H\"ofling and Ryoichi Yamamoto for useful discussions. We 
further thank Felix H\"ofling for sharing unpublished data of the 2D 
Lorentz model with us. 
We thank the EPSRC, the DFG research unit FOR-1394 ``Nonlinear response to probe vitrification'' (HO 2231/7-2), the Royal Society and the ERC (ERC Starting Grant 279541-IMCOLMAT) for financial support.
\end{acknowledgements}
%





\end{document}